\documentclass[]{fairmeta_gr2}

\microtypesetup{expansion=false}

\usepackage{amsmath}
\usepackage{amssymb}
\usepackage{mathtools}
\usepackage{pifont}
\usepackage{wrapfig}
\usepackage{enumitem}

\newlist{tightitemize}{itemize}{1}
\setlist[tightitemize]{leftmargin=*,itemsep=2pt,topsep=2pt,parsep=0pt,label=$\bullet$}

\title{End-to-End Dynamic Sparsity for Resource-Adaptive LLM Inference}

\author[2,*]{Yuhang Chen}
\author[2]{Jinhao Duan}
\author[2]{Ruichen Zhang}
\author[1]{Mingfu Liang}
\author[1]{Xiaohan Wei}
\author[1]{Yunchen Pu}
\author[1]{Fei Tian}
\author[1]{Chonglin Sun}
\author[1]{Parish Aggarwal}
\author[1]{Frank Shyu}
\author[1]{Luke Simon}
\author[1]{Sandeep Pandey}
\author[2,\dagger]{Tianlong Chen}
\author[1,\dagger]{Xi Liu}

\affiliation[1]{Meta AI}
\affiliation[2]{University of North Carolina at Chapel Hill}

\contribution[*]{Work done during an internship at Meta.}
\contribution[\dagger]{Joint corresponding authors.}

\correspondence{Tianlong Chen (\email{tianlong@cs.unc.edu}) and Xi Liu (\email{xliu1@meta.com})}

\abstract{Large Language Models (LLMs) inference is typically deployed under a static resource assumption, where models execute a fixed computational graph regardless of the runtime environment. However, real-world cloud infrastructure is inherently dynamic, characterized by fluctuating availability (e.g., spot instance preemption) and tiered Quality-of-Service requirements. In such volatile settings, static models are inflexible: they either crash under resource constraints or waste compute on redundant operations.
To bridge this gap, we propose \textit{Learning to Allocate} (\texttt{L2A}), an end-to-end framework for resource-adaptive inference. Unlike prior adaptive methods that condition only on input difficulty, we formulate inference as a constrained resource allocation problem that conditions on both the input and the runtime resource budget itself. We introduce lightweight, budget-conditioned and input-aware gating networks integrated into the LLM. These gates are trained via a unified objective that jointly optimizes task performance, logical consistency, and resource costs decoupled along three axes that match how real-world dynamics manifest in deployment: layer skipping for memory and depth pressure, head pruning for throughput contention, and reasoning-token reduction for latency tightening. Crucially, this allows the model to learn a budget-aware policy beyond input difficulty alone: it adaptively configurates its computational footprint with respect to real-time resource dynamics, maximizing reasoning depth for complex tasks when resources permit, while enforcing strict frugality when budgets tighten.
A single \texttt{L2A} model traces the entire compute-accuracy Pareto frontier on Llama-3-8B and Qwen-3-4B: at up to 34\% realized layer sparsity, it stays within 0.6\% of the dense baseline on GSM8K, with the same gap holding zero-shot on out-of-distribution tasks, while every static or heuristic baseline requires a separately tuned model and still drops by 5-10\% at comparable inference time.}

\date{\today}

\begin{document}

\maketitle

\section{Introduction}

Large Language Models (LLMs) have emerged as the cornerstone of modern artificial intelligence~\citep{gpt,instruct-gpt}, demonstrating remarkable capabilities across a wide range of tasks~\citep{llama3,qwen}. However, the deployment of these models in real-world production environments faces a fundamental and increasingly critical challenge: the mismatch between static inference architectures and dynamic resource availability. Current state-of-the-art LLMs are designed as static computational graphs. They operate under a rigid one-size-fits-all paradigm, activating every layer and attention head for every query, regardless of whether the input is a trivial factual question or a complex multi-step reasoning problem. This design assumes a stable, dedicated resource environment, which contradicts the reality of modern cloud computing.

In the real world, LLM inference occurs in dynamic, multi-tenant cloud environments where resources are unpredictable and fluctuate continuously. Available GPU memory and utilization limits can drop dramatically due to competing workloads. This volatility exposes the fragility of static inference systems. When resources shrink, static models face a binary choice: crash due to Out-of-Memory (OOM) errors or stall with unacceptable latency. They lack the elasticity to degrade gracefully. This limitation is particularly acute in two high-value scenarios: \ding{182} Task Survival under Preemption~\citep{lin2022methods}, where models running on low-cost Spot Instances receive a sudden shutdown warning (e.g., 2 minutes remaining) and must race to complete the inference before resources vanish; and \ding{183} Quality of Service (QoS) Differentiation~\citep{li2022managing}, where service providers need to dynamically allocate compute budgets to serve different user tiers (e.g., Premium vs. Free) with varying quality-latency trade-offs. The central thesis of our work is that current LLM systems lack the necessary elasticity to handle these scenarios, leading to brittle and unreliable service.

\begin{wrapfigure}[17]{r}{0.45\linewidth}
\vspace{-0.6em}
\centering
\includegraphics[width=\linewidth]{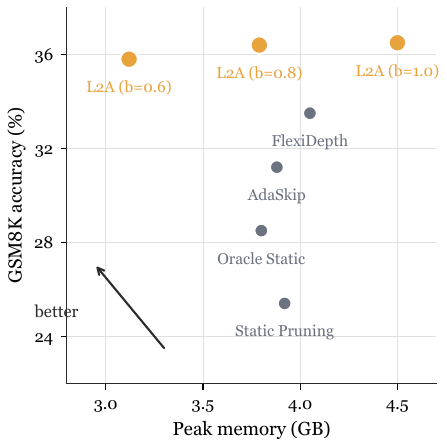}
\caption{Memory--accuracy on Llama-3-8B GSM8K: \texttt{L2A} dominates static and heuristic baselines.}
\label{fig:pareto_teaser}
\end{wrapfigure}

Existing solutions to this efficiency problem fall short of runtime adaptability. Static compression methods like pruning~\citep{ma2023llm,ashkboos2024slicegpt} and distillation~\citep{hsieh2023distilling} produce permanently smaller models that cannot scale up for hard tasks~\citep{shortgpt}. Conversely, dynamic inference techniques, such as early-exit networks~\citep{deebert,fastbert,calm}, often rely on hand-crafted heuristics (e.g., entropy thresholds) that are brittle and struggle to generalize across diverse domains~\citep{layerskip,safe-earlyexit}. A critical open question remains:
\begin{tcolorbox}[before skip=0.2cm, after skip=0.2cm, boxsep=0.0cm, middle=0.1cm, top=0.1cm, bottom=0.1cm]
\textit{How can we enable LLMs to dynamically allocate computational resources at inference time based on the intrinsic difficulty of each specific input and the real-time resource budget?}
\end{tcolorbox}

To bridge this gap, we propose a novel framework for runtime-adaptive inference that unifies dynamic resource allocation with parameter-efficient fine-tuning. Unlike prior methods that treat model compression and adaptation as separate stages, we formulate the problem as a joint optimization objective. We introduce lightweight, input-dependent gating networks into the Transformer architecture, which learn to predict the necessity of each computational block (layers and heads) from the hidden states in real-time. By integrating these gates with Low-Rank Adaptation (LoRA) and a unified loss function that explicitly balances task performance, logical consistency, and multidimensional resource costs, we enable the model to learn an optimal, dynamic inference policy end-to-end.

We validate our framework through extensive experiments on a diverse suite of benchmarks~\citep{eval-harness}, including GSM8K~\citep{gsm8k}, OpenWebText, and HumanEval~\citep{humaneval}. Our results demonstrate that our input-dependent, learned policy achieves a superior performance-efficiency Pareto frontier compared to static or heuristic baselines. Crucially, we show that our method learns a generalizable perception of task difficulty: it automatically reduces computation for simple inputs while preserving full capacity for hard reasoning tasks, even on out-of-distribution domains like code generation.

In conclusion, our core contributions are:
(\textbf{\textit{i}})  We propose a unified optimization framework for dynamic gating networks to learn input-dependent resource allocation policies end-to-end;
(\textbf{\textit{ii}}) We demonstrate through extensive experiments that our method achieves superior elasticity and robustness, effectively handling both resource fluctuations and QoS differentiation requirements.

\section{Related Works}

\noindent\textbf{Efficient and Adaptive LLM Inference.}
Recent work studies conditional computation for Transformer-based LLMs, allocating depth per token/input rather than always executing all layers. Early-exit and adaptive inference methods~\citep{schuster2022confident,liu2020fastbert,deebert} train intermediate predictions to approximate the final output, enabling faster decoding on easy inputs~\citep{calm,layerskip,safe-earlyexit}. Complementary approaches learn to skip computation inside the backbone: Mixture-of-Depths routes a subset of tokens through each layer under a fixed token budget~\citep{raposo2024mixture}, while layer-skipping methods learn lightweight predictors for when layers or submodules can be omitted during inference~\citep{zeng2023learning,fan2024not,luo2025diffskip}~\citep{skiplayer,mod,dlo,mindskip,duollm,liang2026generative}. SkipGPT further highlights that stable, token-aware dynamic pruning benefits from decoupled policies and staged optimization~\citep{zhao2025skipgpt}.

\noindent\textbf{Model Compression under Deployment Constraints.}
A parallel literature compresses LLMs statically for deployment by removing weights or structures (e.g., heads/layers), then fine-tuning to recover accuracy~\citep{ma2023llm,kim2024shortened,shortgpt}~\citep{laco,finercut}. More recent structured approaches compress LLMs by deleting rows/columns in weight matrices or other architecture-aware transformations to preserve throughput on commodity hardware~\citep{ashkboos2024slicegpt}. These static methods are simple to deploy but typically produce a single fixed point on the accuracy--cost curve; meeting diverse runtime budgets often requires multiple models or additional control logic. Our framework complements static compression by learning input-dependent allocation at inference time.

\noindent\textbf{Training Learned Gates.}
Learning to skip components introduces a training--inference gap (soft gates vs. hard decisions) and risks gate collapse. Prior work mitigates these issues via self-distillation and step-wise distillation, encouraging sparse students to match dense teachers and improving robustness of adaptive computation~\citep{hsieh2023distilling,liu2020fastbert,pceebert}. In budgeted or cascaded systems, distillation-like objectives are also used to preserve decision quality while reducing cost~\citep{fanconi2025cascaded}.

\section{Methodology}

\begin{figure}[t]
\centering
\includegraphics[width=\linewidth]{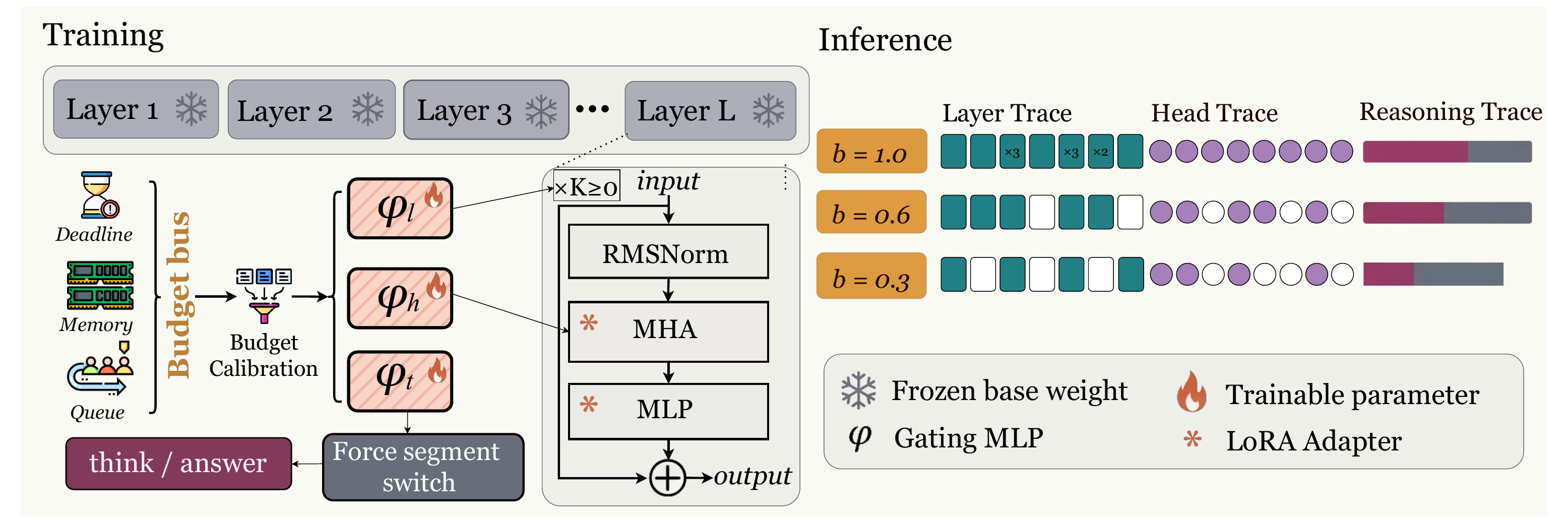}
\caption{\texttt{L2A} framework. The runtime budget $b$ conditions three trainable gates inside a frozen LLM: a layer-skip gate, a head-prune gate, and a reasoning-stop gate. A single trained model adapts its compute footprint to changing resource dynamics at inference time.}
\label{fig:framework}
\end{figure}

In this section, we present our framework for Runtime-Adaptive Inference, which transforms the static computation graph of Large Language Models (LLMs) into a dynamic, input-dependent structure. We formulate the problem as a joint optimization of parameter-efficient adaptation and dynamic resource allocation. By introducing lightweight gating networks and a unified objective function, we enable the model to learn an optimal policy for skipping layers and pruning attention heads on-the-fly, balancing task performance with computational efficiency.

\subsection{Preliminaries and Notations}
\label{sec:method_notations}

We denote the frozen base language model by $f_{\theta}$ and the parameter-efficient adapted model by $f_{\theta+\Delta\theta}$, where $\Delta\theta$ corresponds to the LoRA parameters. Given a tokenized sequence $x$, we denote the next-token distributions of the teacher (dense) and student (budgeted) models by $P_{t}(\cdot\mid x_{<t};\theta)$ and $P_{s}(\cdot\mid x_{<t};\theta,\Delta\theta,\Phi)$, respectively.

We study budget-conditioned dynamic execution in a Transformer with $L$ layers and $H$ attention heads per layer. The input hidden state of layer $l$ is $h_{l-1} \in \mathbb{R}^{d}$. To capture runtime constraints, we introduce a scalar budget signal $b \in [0,1]$ (larger means more compute allowed) and map it to an embedding vector $e(b)$.

Our policy consists of layer gates $\alpha_l(\cdot) \in (0,1)$, head gates $\beta_{l,h}(\cdot) \in (0,1)$, and a budget-conditioned \emph{segment-transition} gate $\tau_t(\cdot) \in (0,1)$. We denote the expected compute by $\mathcal{C}(b)$ and a monotone target budget by $\mathcal{B}(b)$. For training objectives, we use $\mathcal{L}_{\mathrm{ce}}$ for next-token cross-entropy, $\mathcal{L}_{\mathrm{kd}}$ for KL distillation, and $\mathcal{L}(b)$ for the total budget-conditioned loss. We use $\mathcal{S}$ to denote a mini-batch of inputs.

To make the transition behavior explicit, we assume a structured generation format
\texttt{<think>}$\cdots$\texttt{</think>}
\texttt{<answer>}$\cdots$\texttt{</answer>}. During training, we insert these tags into the target sequences so the model learns the segmentation. At inference time, the transition gate can force emitting the boundary token sequence \texttt{</think><answer>} to stop generating further reasoning tokens and enter the answer segment, rather than prematurely ending the whole sequence.

\subsection{Learnable Dynamic Allocation Mechanism}

To achieve fine-grained control over computational resources, we introduce learnable gating modules into the pre-trained LLM backbone. Crucially, our gates are \emph{budget-conditioned}: in addition to hidden states, they take an external budget signal $b$ that specifies the maximum allowed compute.

\noindent\textbf{Budget signal.}
We represent deployment constraints with a scalar budget $b \in [0,1]$ (larger means more compute allowed) and feed it to all gating networks through a learned embedding $e(b)$. During training, we sample $b \sim \mathcal{D}$ to expose the policy to diverse operating regimes (we later ablate different choices of $\mathcal{D}$). At inference time, $b$ is obtained by calibrating real-time system signals (e.g., deadline, queue length, memory headroom) into the same normalized range.

\noindent\textbf{Dynamic Layer Skipping.}
For a Transformer layer $l$, we introduce a lightweight gating network $\phi_L^{(l)}$ that produces a budget-conditioned scalar $\alpha_l \in (0,1)$:
\begin{equation}
\label{eq:layer_gate}
    \alpha_l(h_{l-1}, b) = \sigma\big(\text{MLP}_{\phi_L}^{(l)}([h_{l-1}; e(b)])\big),
\end{equation}
where $\sigma$ is the sigmoid and $[\cdot;\cdot]$ is concatenation. The layer's forward pass becomes a soft-gated residual $h_l = h_{l-1} + \alpha_l \cdot \mathcal{F}_l(h_{l-1})$, with $\mathcal{F}_l$ the standard self-attention$+$MLP sub-layer mapping. During training $\alpha_l$ scales the residual update so gradients flow into $\phi_L$; we optionally anneal the gate temperature to push $\alpha_l$ near-binary by end of training. At inference, a hard threshold (e.g., $0.5$) binarizes $\alpha_l$, physically skipping $\mathcal{F}_l$ when $\alpha_l$ falls below it.

\noindent\textbf{Dynamic Head Pruning.}
Head gates $\beta_{l,h} \in (0,1)$ for each head $h$ in layer $l$ are produced by a network $\phi_H^{(l,h)}$ of the same form as Eq.~\ref{eq:layer_gate}, conditioned on $(h_{l-1}, b)$. The MHA output becomes a $\beta$-weighted concatenation $\text{MHA}(h_{l-1}) = [\beta_{l,1}\,\text{Head}_1, \dots, \beta_{l,H}\,\text{Head}_H]\,W_O$, suppressing less relevant heads based on input context and budget. The wall-clock benefit from head pruning is implementation-dependent; our reported speedups come primarily from layer skipping and reasoning-length reduction.

\noindent\textbf{Dynamic Reasoning-to-Answer Transition.}
Long reasoning traces often dominate inference latency on complex tasks. We add a transition gate $\tau_t \in (0,1)$, produced by $\phi_T$ from $(h_t, b)$ at each generation step (same form as Eq.~\ref{eq:layer_gate}). During training, $\tau_t$ remains soft and contributes to a reasoning-length cost (below); at inference, when $\tau_t$ exceeds a threshold, we force-emit the boundary tokens \texttt{</think><answer>} (\Cref{sec:method_notations}), after which decoding continues in the answer segment.

\subsection{Parameter-Efficient Adaptation}

Directly skipping layers or heads in a pre-trained model can severely disrupt its internal representations. To mitigate this and enable the model to adapt to sparse execution paths, we employ Low-Rank Adaptation (LoRA). We freeze the original model parameters $W_{0}$ and inject trainable low-rank matrices $\Delta W = BA$ into the linear layers (Query, Key, Value, Output, etc.). The effective weight becomes $W = W_{0} + \Delta W$.
This ensures that the optimization process is parameter-efficient and does not catastrophic forget the pre-trained knowledge, while allowing the backbone to adjust to the dynamic gating behavior.

\subsection{Unified Optimization Objective}

We define a unified loss function that jointly optimizes the LoRA parameters ($W_{PEFT}$) and the gating network parameters ($\Phi = \{\phi_L, \phi_H, \phi_T\}$). The objective balances three competing goals: maximizing task accuracy, preserving the original model's behavior, and satisfying a runtime budget.

For each training instance, we sample a budget $b \sim \mathcal{D}$ and condition all gates on $b$. We then optimize a budget-constrained objective by penalizing deviations between the expected compute $\mathcal{C}(b)$ and the target budget $\mathcal{B}(b)$:
\begin{equation}
\begin{split}
    \mathcal{L}(b) = \mathcal{L}_{\mathrm{ce}} + \gamma \cdot \mathcal{L}_{\mathrm{kd}} + \lambda \cdot \big(\mathcal{C}(b) - \mathcal{B}(b)\big) + \lambda_T \cdot \mathcal{R}_{\mathrm{tok}}
\end{split}
\end{equation}
where $\lambda$ controls the strength of budget enforcement and can be interpreted as a Lagrange multiplier.

\noindent\textbf{Competence preservation and distillation.}
$\mathcal{L}_{\mathrm{ce}} = -\sum_t \log P_s(y_t \mid x_{<t};\theta,\Delta\theta,\Phi)$ is the standard next-token cross-entropy and preserves generative capability under sparsity. $\mathcal{L}_{\mathrm{kd}} = \sum_t D_{KL}(P_t \,\|\, P_s)$ is the forward-KL distillation from the frozen dense teacher and prevents the student from drifting when computation is skipped under tight budgets.

\noindent\textbf{Budgeted compute cost.}
We estimate the expected compute as $\mathcal{C}(b) = \mathcal{R}_L(b) + \mathcal{R}_H(b)$, with
\begin{equation}
    \mathcal{R}_L(b) = \tfrac{1}{|\mathcal{S}| L} \!\!\sum_{x \in \mathcal{S}}\sum_{l=1}^{L} \alpha_l(x,b)\,C_l, \quad
    \mathcal{R}_H(b) = \tfrac{1}{|\mathcal{S}| L H} \!\!\sum_{x \in \mathcal{S}}\sum_{l,h} \beta_{l,h}(x,b)\,C_h,
\end{equation}
where $C_l, C_h$ are normalized per-layer / per-head FLOPs or latency costs. The target $\mathcal{B}(b)$ is a monotone function of $b$ so larger $b$ permits more compute.

\noindent\textbf{Reasoning length cost.}
Rather than penalizing per-component activation, here we penalize the expected length of the \texttt{<think>} segment via $\mathcal{R}_{\mathrm{tok}} = \tfrac{1}{|\mathcal{S}|} \sum_x \sum_{t=1}^{T} (1 - \tau_t(x))$. Minimizing $\mathcal{R}_{\mathrm{tok}}$ pushes $\tau_t \to 1$ at earlier decoding steps, shortening reasoning under tight budgets while $\mathcal{L}_{\mathrm{ce}}$ keeps the final answer correct.

\subsection{Training and Inference}

The training process involves an end-to-end joint optimization on a mixed dataset containing both simple and complex samples. For each sample, we additionally draw a budget $b \sim \mathcal{D}$ and condition the gates on $b$. The gradients from $\mathcal{L}(b)$ update both the LoRA parameters (to adapt to sparse execution paths) and the gating parameters (to learn a budget-aware routing policy). Intuitively, the budget term encourages the model to reduce compute under tight $b$, while $\mathcal{L}_{\mathrm{ce}}$ and $\mathcal{L}_{\mathrm{kd}}$ preserve quality.

At inference time, we compute the budget-conditioned gates and replace soft gating with hard decisions. For layers, we skip the layer entirely if $\alpha_l \leq \tau$, achieving real-world acceleration without any specialized hardware support.

\begin{table}[!t]
\centering
\small
\caption{Metric definitions used throughout the experiments.}
\label{tab:metric_definitions}
\begin{tabular}{ll}
\toprule
\textbf{Metric} & \textbf{Definition} \\
\midrule
Avg. Sparsity (L/H) & Mean fraction of skipped layers / pruned heads across the evaluation set. \\
Avg. Mem (GB) & Peak GPU memory during inference, averaged across batches. \\
Time (s) & Total wall-clock inference time over the evaluation set under the same pipeline. \\
\bottomrule
\end{tabular}
\end{table}

\begin{table}[!t]
\centering
\small
\caption{Baseline comparison on Llama-3-8B at representative operating points. Higher is better for Acc/Score/Pass@1 and lower is better for PPL, memory, and time. Avg. Sparsity (L/H) reports the average realized layer/head sparsity over the evaluation set.}
\label{tab:baselines_llama}
\resizebox{\linewidth}{!}{%
\begin{tabular}{lccccccccc}
\toprule
\textbf{Method} & \textbf{\shortstack{OWT\\PPL $\downarrow$}} & \textbf{\shortstack{GSM8K\\Acc $\uparrow$}} & \textbf{\shortstack{MMLU\\Acc $\uparrow$}} & \textbf{\shortstack{Alpaca\\Score $\uparrow$}} & \textbf{\shortstack{HumanEval\\Pass@1 $\uparrow$}} & \textbf{\shortstack{BBH\\Acc $\uparrow$}} & \textbf{\shortstack{Avg. Sparsity\\(L/H)}} & \textbf{\shortstack{Mem\\GB $\downarrow$}} & \textbf{\shortstack{Time\\s $\downarrow$}} \\
\midrule
Original Model & 12.45 & 36.5\% & 32.2\% & 45.0 & 12.8\% & 30.5\% & 0\% / 0\% & 4.50 & 3452 \\
Static Pruning & 13.8 & 25.4\% & 24.1\% & 40.5 & 5.2\% & 21.3\% & 33\% / 0\% & 3.92 & 2310 \\
Oracle Static Pruning & \textbf{12.55} & 28.5\% & 27.8\% & 42.1 & 7.8\% & 24.5\% & 33\% / 20\% & \textbf{3.80} & \textbf{2250} \\
AdaSkip & 12.65 & 31.2\% & 28.5\% & 43.0 & 9.5\% & 26.8\% & 35\% / -- & 3.88 & 2350 \\
FlexiDepth & 12.6 & 33.5\% & 29.8\% & 43.8 & 10.8\% & 28.2\% & 32\% / -- & 4.05 & 2420 \\
Ours & 12.58 & \textbf{35.9\%} & \textbf{31.5\%} & \textbf{44.2} & \textbf{12.2\%} & \textbf{29.8\%} & 34\% / 24\% & 4.10 & 2380 \\
\bottomrule
\end{tabular}%
}
\end{table}
\begin{table}[!t]
\centering
\small
\caption{Baseline comparison on Qwen-3-4B at representative operating points. Higher is better for Acc/Score/Pass@1 and lower is better for PPL, memory, and time. Avg. Sparsity (L/H) reports the average realized layer/head sparsity over the evaluation set.}
\label{tab:baselines_qwen}
\resizebox{\linewidth}{!}{%
\begin{tabular}{lccccccccc}
\toprule
\textbf{Method} & \textbf{\shortstack{OWT\\PPL $\downarrow$}} & \textbf{\shortstack{GSM8K\\Acc $\uparrow$}} & \textbf{\shortstack{MMLU\\Acc $\uparrow$}} & \textbf{\shortstack{Alpaca\\Score $\uparrow$}} & \textbf{\shortstack{HumanEval\\Pass@1 $\uparrow$}} & \textbf{\shortstack{BBH\\Acc $\uparrow$}} & \textbf{\shortstack{Avg. Sparsity\\(L/H)}} & \textbf{\shortstack{Mem\\GB $\downarrow$}} & \textbf{\shortstack{Time\\s $\downarrow$}} \\
\midrule
Original Model & 11.20 & 48.50\% & 45.10\% & 58.0 & 22.40\% & 41.20\% & 0\% / 0\% & 5.80 & 4100 \\
Static Pruning & 12.95 & 32.80\% & 34.50\% & 51.2 & 8.50\% & 28.40\% & 33\% / 0\% & 5.10 & 2750 \\
Oracle Static Pruning & 11.35 & 36.20\% & 39.80\% & 53.5 & 13.10\% & 33.50\% & 33\% / 22\% & \textbf{4.95} & \textbf{2680} \\
AdaSkip & 11.42 & 40.50\% & 41.20\% & 54.8 & 16.20\% & 36.10\% & 36\% / -- & 5.65 & 2800 \\
FlexiDepth & 11.38 & 44.10\% & 42.50\% & 56.1 & 18.80\% & 38.40\% & 34\% / -- & 5.75 & 2910 \\
Ours & \textbf{11.32} & \textbf{47.80\%} & \textbf{44.50\%} & \textbf{57.2} & \textbf{21.50\%} & \textbf{40.50\%} & \textbf{38\% / 28\%} & 5.30 & 2850 \\
\bottomrule
\end{tabular}%
}
\end{table}

\subsection{Inference-Time Budget Calibration}
\label{sec:budget_calibration}

At deployment, $b$ is produced by calibrating real-time system constraints into the normalized range $[0,1]$ to match the training-time support of $b \sim \mathcal{D}$. We map multiple signals to intermediate budgets and take a conservative aggregate $b = \min\{b_{\mathrm{time}}, b_{\mathrm{mem}}, b_{\mathrm{queue}}\}$. Given a request-level deadline $T_{\mathrm{sla}}$ and an online estimate of the dense-model latency $\widehat{T}_{\mathrm{dense}}(x)$, we set $b_{\mathrm{time}} = \mathrm{clip}(T_{\mathrm{sla}} / \widehat{T}_{\mathrm{dense}}(x), 0, 1)$, so tighter deadlines yield smaller budgets. When running on spot instances with remaining time-to-interruption $T_{\mathrm{spot}}$, we further set $b_{\mathrm{spot}} = \mathrm{clip}(T_{\mathrm{spot}} / \widehat{T}_{\mathrm{dense}}(x), 0, 1)$ and update $b_{\mathrm{time}} \leftarrow \min(b_{\mathrm{time}}, b_{\mathrm{spot}})$. Memory and queue budgets $b_{\mathrm{mem}}, b_{\mathrm{queue}}$ are optionally computed from KV-cache headroom and queue length / utilization, clipped to $[0,1]$, and combined via the same minimum.

\section{Experiments}

We design our experiments to rigorously validate two central hypotheses: (1) Our unified optimization framework with dynamic gating, achieves a superior performance-efficiency trade-off compared to static or heuristic methods. (2) The learned gating policy captures input-dependent difficulty features that generalize across tasks, enabling robust adaptive inference even on out-of-distribution (OOD) data.

\subsection{Experimental Setup}

\noindent\textbf{Base models.} We evaluate on two representative open-source backbones, Llama-3-8B~\citep{llama3} and Qwen-3-4B~\citep{qwen}, to test whether the proposed gating-and-adaptation framework transfers across model families.

\noindent\textbf{Datasets and evaluation split.} We benchmark both in-distribution (ID) performance and out-of-distribution (OOD) robustness.
\begin{tightitemize}
    \item \textbf{ID training/calibration domains}: OpenWebText~\citep{Gokaslan2019OpenWeb} (OWT; general language modeling) and GSM8K~\citep{gsm8k} (math reasoning, metric: accuracy). We additionally report results on MMLU~\citep{mmlu} (knowledge/QA; metric: accuracy) and Alpaca-Eval~\citep{alpaca_eval} (instruction following; metric: score).
    \item \textbf{OOD domains (held out during training)}: HumanEval~\citep{humaneval} (Python code generation; metric: Pass@1) and Big-Bench Hard~\citep{bbh} (BBH; symbolic reasoning; metric: accuracy).
\end{tightitemize}

\noindent\textbf{Decoding and prompting.} All methods share the same prompt template, max-generation length, and stopping criteria; structured-format runs adopt \texttt{<think>}$\cdots$\texttt{</think><answer>}$\cdots$\texttt{</answer>} uniformly to isolate dynamic allocation from prompt engineering.

\noindent\textbf{Baselines.} We compare against (i) \emph{Original}, the full dense model; (ii) \emph{Uniform pruning}, every $k$-th layer skipped irrespective of input; (iii) \emph{Oracle static pruning}, layers / heads removed via offline profiling on ID calibration data; (iv) \emph{AdaSkip}~\citep{he2025adaskip}, an entropy-based early-exit heuristic; and (v) \emph{FlexiDepth}~\citep{luo2025adaptive}, learned adaptors with an auxiliary skipping loss.

\noindent\textbf{Metrics.} Quality: GSM8K / MMLU / BBH accuracy, Alpaca-Eval score, HumanEval Pass@1, and OWT perplexity. Efficiency: average realized layer / head sparsity, peak GPU memory, and total inference time. Definitions of efficiency metrics are in Table~\ref{tab:metric_definitions}.

\noindent\textbf{Implementation details.} LoRA is applied to all linear layers at rank $r{=}16$; the gate networks are 2-layer MLPs; we use a temperature-annealing schedule for the continuous gate relaxation~\citep{mixer} and switch to hard gating at inference. For representative operating points (Tables~\ref{tab:baselines_llama}--\ref{tab:baselines_qwen}), each baseline is tuned to the same efficiency regime as our method (matched average time / sparsity), and the evaluation pipeline (batching, decoding, hardware) is held identical across methods.

\subsection{Main Results}

We present the primary comparison of our method against all baselines in terms of the accuracy-efficiency Pareto frontier. The results demonstrate that our method consistently achieves the higher accuracy for any given computational budget. Specifically, under dynamic resource constraint, our method maintains near-lossless performance on GSM8K, whereas static uniform pruning suffers catastrophic degradation. While Oracle Static Pruning performs well on simple texts (OWT), it fails to adapt to the reasoning demands of GSM8K. Similarly, while AdaSkip offers some dynamic benefits, its reliance on entropy as a proxy for difficulty proves less robust than our learned difficulty features. This validates the effectiveness of our unified formulation in finding the optimal dynamic compression strategy.

\begin{figure}[t]
\centering
\begin{subfigure}[t]{0.32\linewidth}
  \centering
  \includegraphics[width=\linewidth]{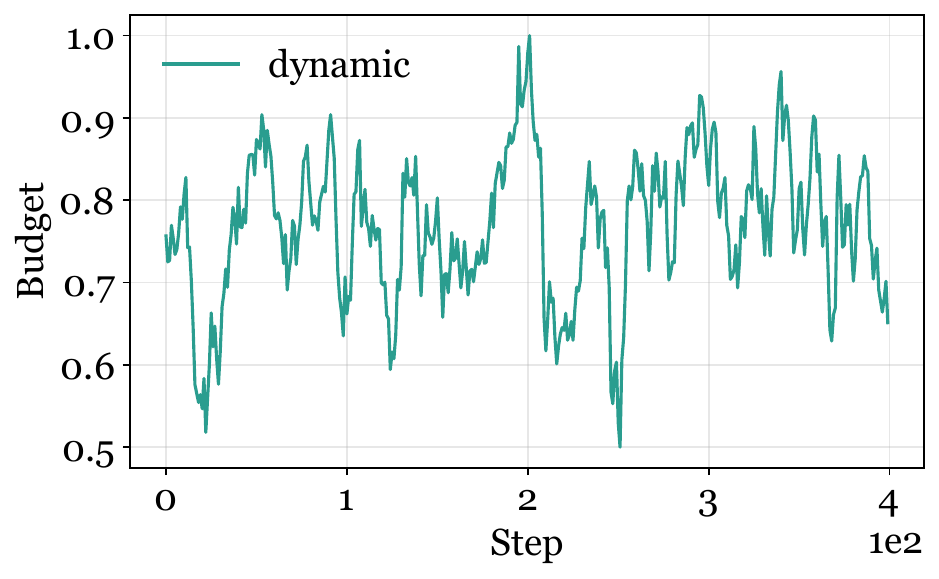}
  \caption{Budget schedule and allocation.}
\end{subfigure}\hfill
\begin{subfigure}[t]{0.32\linewidth}
  \centering
  \includegraphics[width=\linewidth]{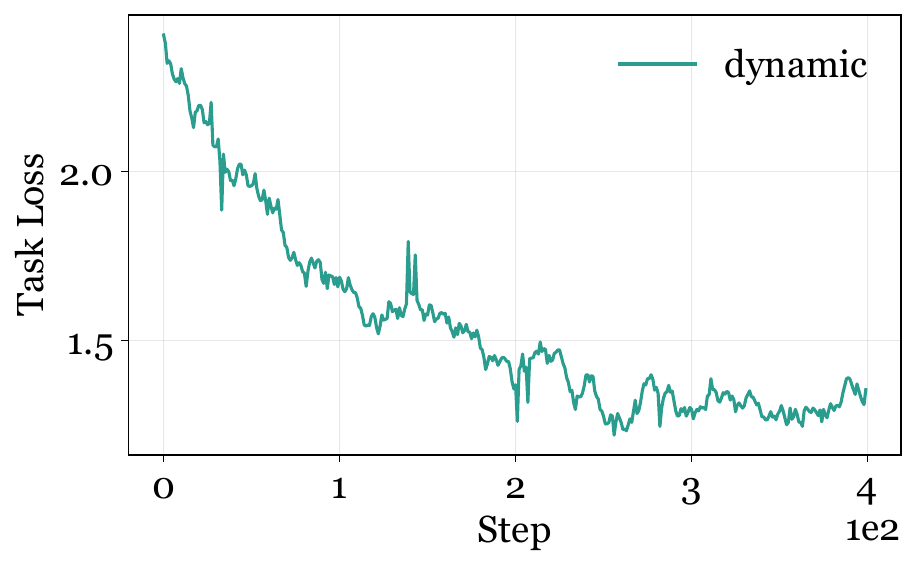}
  \caption{Training task loss.}
\end{subfigure}\hfill
\begin{subfigure}[t]{0.32\linewidth}
  \centering
  \includegraphics[width=\linewidth]{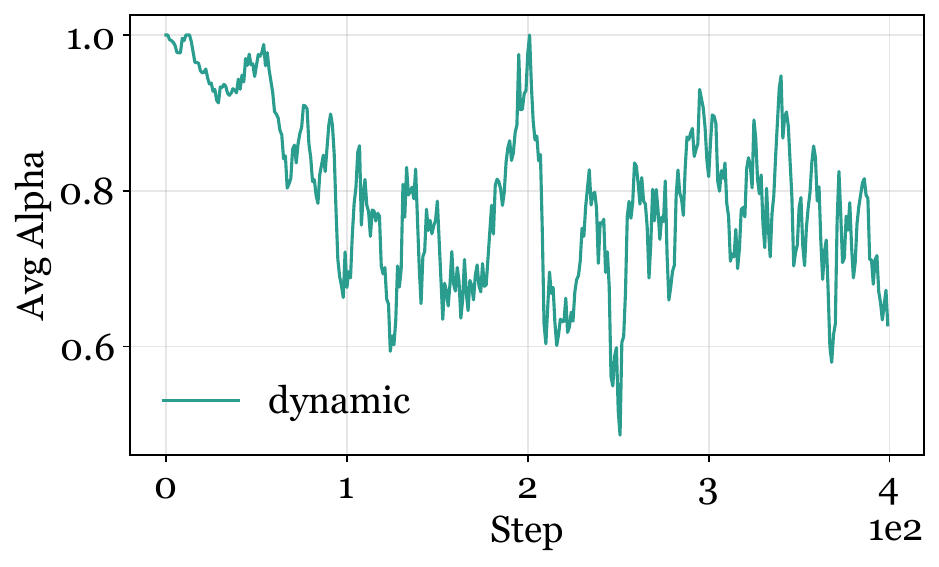}
  \caption{Mean layer gate $\alpha$.}
\end{subfigure}
\caption{Dynamic-budget diagnostics. (a) shows the time-varying budget schedule and the corresponding realized allocation behavior; (b) reports the training task loss when the policy is conditioned on a changing budget; and (c) reports the evolution of the mean layer gate $\alpha$, illustrating that the learned policy adjusts its retention level in response to budget changes.}
\label{fig:dynamic_budget_diagnostics}
\end{figure}

\noindent\textbf{Baseline Comparison.}
Table~\ref{tab:baselines_llama} and Table~\ref{tab:baselines_qwen} summarize representative operating points on Llama-3-8B and Qwen-3-4B.
Across both backbones, static pruning methods achieve meaningful speedups but suffer larger performance drops, especially on reasoning-intensive and OOD tasks, because they apply a fixed architecture to all inputs.
In contrast, adaptive baselines (AdaSkip/FlexiDepth) better preserve performance under similar compute, and our method further closes the gap to the dense model while maintaining comparable latency.
As shown in Figure~\ref{fig:dynamic_budget_diagnostics}, the learned policy tracks the externally provided budget signal by adjusting both the realized allocation and the average layer retention, while maintaining stable optimization dynamics.

\begin{table}[t]
\centering
\small
\caption{Fixed-budget evaluation on Llama-3-8B. For each target budget (relative to original compute), we compare a naive baseline (FixedBudget), a \emph{Static Oracle} (best fixed configuration found via search), and our dynamic policy matched to the same average compute cost.}
\label{tab:fixedbudget_llama}
\resizebox{\linewidth}{!}{%
\begin{tabular}{llccccccccc}
\toprule
\textbf{Budget} & \textbf{Method} & \textbf{\shortstack{OWT\\PPL $\downarrow$}} & \textbf{\shortstack{GSM8K\\Acc $\uparrow$}} & \textbf{\shortstack{MMLU\\Acc $\uparrow$}} & \textbf{\shortstack{Alpaca\\Score $\uparrow$}} & \textbf{\shortstack{HumanEval\\Pass@1 $\uparrow$}} & \textbf{\shortstack{BBH\\Acc $\uparrow$}} & \textbf{\shortstack{Sparsity\\(L/H)}} & \textbf{\shortstack{Mem\\GB $\downarrow$}} & \textbf{\shortstack{Time\\s $\downarrow$}} \\
\midrule
1.0 & Original & 12.45 & 36.5\% & 32.2\% & 45.0 & 12.8\% & 30.5\% & 0.0\% / 0.0\% & 4.50 & 3452 \\
\midrule
\multirow{3}{*}{0.8} & FixedBudget & 13.16 & 33.6\% & 30.8\% & 44.6 & 11.0\% & 29.2\% & 20.0\% / 20.0\% & 3.82 & 2872 \\
& Static Oracle & 12.65 & 35.2\% & 31.8\% & 44.9 & 12.1\% & 29.8\% & 20.0\% / 20.0\% & 3.80 & 2850 \\
& Ours & \textbf{12.50} & \textbf{36.4\%} & \textbf{32.1\%} & \textbf{45.2} & \textbf{12.7\%} & \textbf{30.4\%} & 24.1\% / 15.7\% & \textbf{3.79} & \textbf{2845} \\
\midrule
\multirow{3}{*}{0.6} & FixedBudget & 13.70 & 31.7\% & 28.8\% & 43.9 & 9.3\% & 26.7\% & 40.0\% / 40.0\% & 3.15 & 2320 \\
& Static Oracle & 12.88 & 33.5\% & 30.2\% & 44.3 & 10.5\% & 28.1\% & 40.0\% / 40.0\% & \textbf{3.10} & \textbf{2290} \\
& Ours & \textbf{12.58} & \textbf{35.8\%} & \textbf{31.6\%} & \textbf{44.8} & \textbf{12.0\%} & \textbf{29.9\%} & 47.1\% / 30.8\% & 3.12 & 2305 \\
\bottomrule
\end{tabular}
}
\end{table}

\noindent\textbf{Fixed-Budget Evaluation.}
To distinguish \emph{static} resource constraints from \emph{dynamic} resource-adaptive inference, we additionally evaluate a FixedBudget baseline, where a single global sparsity level is enforced for all inputs (i.e., the model cannot allocate more compute to hard examples). Table~\ref{tab:fixedbudget_llama} reports results on Llama-3-8B under two budgets.

As the fixed budget tightens (higher global sparsity), we observe a consistent degradation across both ID and OOD tasks, reflecting the fundamental limitation of static allocation: it must under-compute on hard samples to satisfy the same budget that is appropriate for easy samples. This motivates our dynamic policy, which can preserve capacity on logically demanding inputs while still saving compute on trivial ones.

\subsection{Cross-Task Generalization and Robustness}

This section addresses the critical question of whether the learned policy simply overfits to the training data or learns a generalizable perception of difficulty. We conduct a Super Matrix cross-evaluation experiment, where models trained on single domains are tested against unseen domains.

\begin{table}[t]
\centering
\small
\caption{Super-matrix cross-domain evaluation on Llama-3-8B. Each row trains (or calibrates) the policy on a \emph{single} source domain and is evaluated on a shared multi-benchmark suite. We report task performance along with realized average sparsity (Layer/Head), peak GPU memory, and total inference time. This experiment highlights how single-domain training can lead to over-pruning on hard tasks or under-pruning on easy tasks.}
\label{tab:supermatrix_llama}
\resizebox{\linewidth}{!}{%
\begin{tabular}{lccccccccc}
\toprule
\textbf{Training Source} & \textbf{\shortstack{OWT\\PPL $\downarrow$}} & \textbf{\shortstack{GSM8K\\Acc $\uparrow$}} & \textbf{\shortstack{MMLU\\Acc $\uparrow$}} & \textbf{\shortstack{Alpaca\\Score $\uparrow$}} & \textbf{\shortstack{HumanEval\\Pass@1 $\uparrow$}} & \textbf{\shortstack{BBH\\Acc $\uparrow$}} & \textbf{\shortstack{Avg. Sparsity\\(L/H)}} & \textbf{\shortstack{Mem\\GB $\downarrow$}} & \textbf{\shortstack{Time\\s $\downarrow$}} \\
\midrule
Original & 12.45 & 36.5\% & 32.2\% & 45.0 & 12.8\% & 30.5\% & 0\% / 0\% & 4.50 & 3452 \\
Fine-tuned on OWT & 12.52 & 34.8\% & 30.5\% & 44.2 & 6.5\% & 29.1\% & 55.2\% / 40.5\% & 3.85 & 2206 \\
Fine-tuned on GSM8K & 12.48 & 36.8\% & 33.8\% & 46.5 & 13.5\% & 30.9\% & 10.4\% / 5.2\% & 4.42 & 3013 \\
Fine-tuned on MMLU & 12.50 & 33.5\% & 31.9\% & 45.8 & 13.2\% & 30.4\% & 25.6\% / 15.3\% & 4.25 & 2632 \\
Fine-tuned on Alpaca & 12.55 & 34.1\% & 30.5\% & 44.8 & 12.8\% & 30.5\% & 30.8\% / 20.1\% & 4.18 & 2594 \\
\bottomrule
\end{tabular}%
}
\end{table}

\begin{table}[!t]
\centering
\small
\caption{Ablation study. We evaluate the contribution of each objective term. \textbf{w/o KD} shows a catastrophic drop in reasoning capabilities, validating the need for logical consistency. \textbf{w/o budget loss} results in a dense model with no acceleration. \textbf{w/o head gating} (layer-only) underperforms the joint strategy at similar latency constraints.}
\label{tab:ablation_main}
\resizebox{\linewidth}{!}{%
\begin{tabular}{lcccccccc}
\toprule
\textbf{Variant} & \textbf{\shortstack{OWT\\PPL $\downarrow$}} & \textbf{\shortstack{GSM8K\\Acc $\uparrow$}} & \textbf{\shortstack{MMLU\\Acc $\uparrow$}} & \textbf{\shortstack{HumanEval\\Pass@1 $\uparrow$}} & \textbf{\shortstack{BBH\\Acc $\uparrow$}} & \textbf{\shortstack{Avg. Sparsity\\(L/H)}} & \textbf{\shortstack{Mem\\GB $\downarrow$}} & \textbf{\shortstack{Time\\s $\downarrow$}} \\
\midrule
Full (Ours) & \textbf{12.6} & \textbf{35.8\%} & \textbf{31.6\%} & \textbf{12.0\%} & \textbf{29.9\%} & 47.1\% / 30.8\% & \textbf{3.1} & \textbf{2305} \\
\midrule
w/o KD ($\gamma=0$) & 21.9 & 14.4\% & 20.5\% & 2.1\% & 11.2\% & 48.5\% / 32.1\% & 3.0 & 2280 \\
w/o budget loss ($\lambda=0$) & 12.4 & 36.5\% & 32.1\% & 12.9\% & 30.6\% & 0.2\% / 0.5\% & 4.5 & 3440 \\
w/o token cost ($\lambda_T=0$) & 12.6 & 36.0\% & 31.7\% & 12.1\% & 30.1\% & 46.5\% / 30.2\% & 3.2 & 2650 \\
w/o head gating ($\beta\equiv 1$) & 12.8 & 33.1\% & 30.2\% & 9.8\% & 27.4\% & 62.5\% / 0.0\% & 3.1 & 2315 \\
\bottomrule
\end{tabular}%
}
\end{table}

\begin{figure}[t]
\centering
\hfill
\begin{subfigure}[t]{0.42\linewidth}
  \centering
  \includegraphics[width=\linewidth]{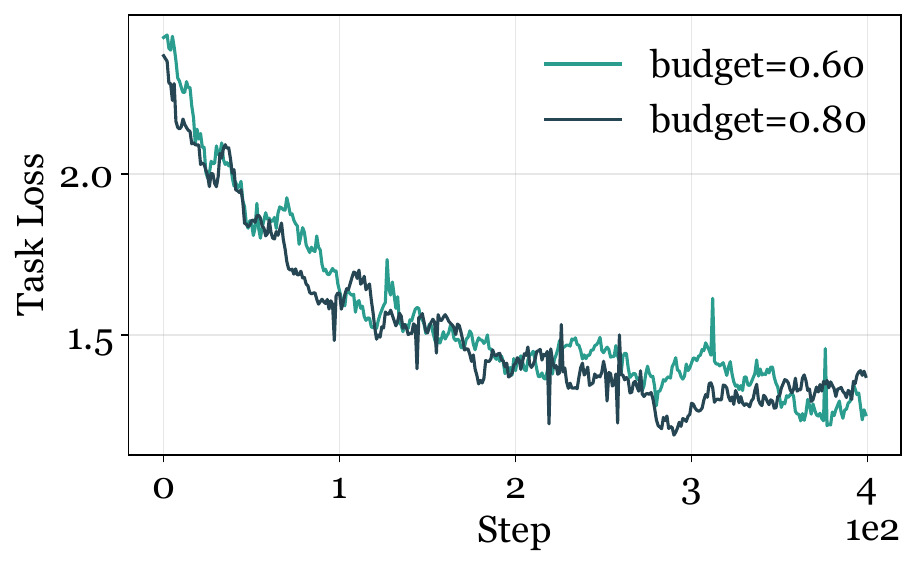}
  \vspace{-15pt}
  \caption{Task loss.}
\end{subfigure}\hfill
\begin{subfigure}[t]{0.42\linewidth}
  \centering
  \includegraphics[width=\linewidth]{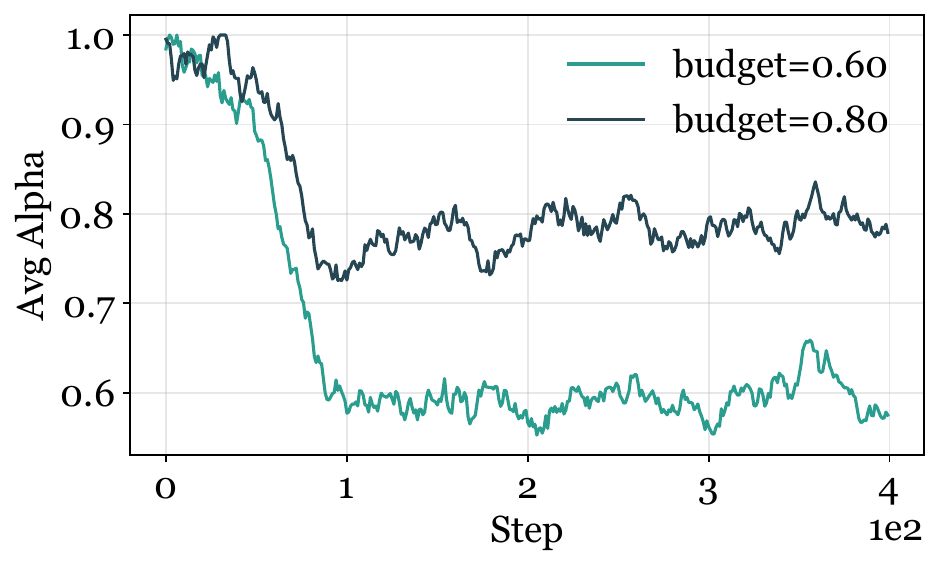}
  \vspace{-15pt}
  \caption{Mean layer gate $\alpha$.}
\end{subfigure}\hfill
\vspace{-5pt}
\caption{Fixed-budget training diagnostics. Under a fixed budget condition, we track (a) the task loss and (b) the mean layer gate $\alpha$ across optimization steps to verify stable convergence and non-collapsing gating behavior.}
\label{fig:fixed_budget_training_curves}
\end{figure}

As shown in Figure~\ref{fig:fixed_budget_training_curves}, the policy converges without gate collapse under a fixed budget, and the mean layer gate stabilizes after the initial adaptation phase.

Table~\ref{tab:supermatrix_llama} reveals a stark asymmetric failure mode in baselines. Models fine-tuned exclusively on easy data (OWT) reach high realized sparsity ($\sim$55\%) but collapse on code generation (HumanEval Pass@1 $12.8\% \to 6.5\%$), reflecting over-pruning on out-of-distribution hard inputs. Models fine-tuned exclusively on hard data (GSM8K) preserve task quality but fail to save compute on easy or instruction-style inputs (Sparsity $\sim$10\%), acting essentially as dense models. By contrast, training across mixed sources lets the policy modulate its sparsity by input domain: it stays conservative on hard reasoning while applying higher sparsity ($\sim$30\%) on simple instruction following (Alpaca-Eval). This supports our hypothesis that the gating networks learn input-difficulty features that transfer across domains rather than memorizing domain-specific patterns.

\subsection{Mechanism Analysis}

We provide quantitative diagnostics to interpret the decision-making process and training stability of the learned gating networks.
In particular, Figure~\ref{fig:dynamic_budget_diagnostics} shows how the realized allocation behavior (and the mean gate $\alpha$) tracks a time-varying budget schedule during training, while Figure~\ref{fig:fixed_budget_training_curves} reports stable convergence of task loss and non-collapsing gate behavior under a fixed-budget setting.

Figure~\ref{fig:dynamic_budget_diagnostics} shows that the policy tracks the non-stationary budget schedule by co-varying its mean gate $\alpha$, while still optimizing stably (loss decreases). Figure~\ref{fig:fixed_budget_training_curves} shows similar stability under fixed budgets, where tighter budgets converge to lower average $\alpha$ (more skipping) than looser budgets.

\subsection{Ablation Studies}

We validate the necessity of each component in our framework as shown in Table~\ref{tab:ablation_main}. Removing distillation ($\gamma=0$) causes a catastrophic collapse in reasoning capabilities (GSM8K: $35.8\% \to 14.4\%$), confirming its critical role in preserving logical consistency under aggressive pruning. Disabling the resource penalty ($\lambda=0$) results in a dense model with zero acceleration, proving that sparsity is explicitly driven by our objective. Furthermore, relying solely on layer skipping ($\beta \equiv 1$) is suboptimal; to match the same latency, it forces excessive depth reduction ($62.5\%$), degrading accuracy compared to our strategy ($47.1\%/30.8\%$), which achieves a superior Pareto trade-off by distributing sparsity across both depth and width.

\subsection{Discussion and Limitations}

Our results should be interpreted in light of several practical considerations. First, the wall-clock benefit from head pruning can be implementation-dependent, whereas layer skipping and reasoning-length control yield more consistent acceleration. Second, the budget calibration mechanism in \Cref{sec:budget_calibration} assumes access to a reasonable online latency/memory estimator; errors in these estimates can lead to conservative or aggressive budgeting. Finally, the structured \texttt{<think>}/\texttt{<answer>} format provides explicit control over reasoning length, but requires consistent prompting and may not be equally applicable to all deployment settings.

\section{Conclusion}
We introduced \texttt{L2A}, a budget-conditioned framework that equips large language models with lightweight gating networks to dynamically allocate computation under runtime constraints. By jointly optimizing task loss, distillation, and budget-matching objectives, our approach can skip layers, suppress attention heads, and shorten the thinking segment while preserving the final answer quality. Experiments across multiple backbones and benchmarks demonstrate improved performance and efficiency trade-offs and stronger robustness than static or heuristic baselines.

\bibliographystyle{assets/plainnat}
\bibliography{example_paper}

\end{document}